%% file: main.tex
\documentclass[aps,pra,reprint,floats,floatfix,amsmath,superscriptaddress,nofootinbib]{revtex4-1}
\usepackage[utf8]{inputenc}
\usepackage{todonotes}
\usepackage[separate-uncertainty=false,per-mode=symbol,inter-unit-product=\cdot]{siunitx}
\usepackage{braket}
\usepackage{hyperref}

\newcommand{\hcf}{HCPBGF}
\newcommand{\gTwo}[1]{$g^{(2)}_{#1}$}

\DeclareSIUnit\pairs{pairs}
\newcommand{\bright}{\pairs\per\second\per\mega\hertz}
\newcommand{\brightMilli}{\pairs\per\second\per\mega\hertz\per\milli\watt}
\newcommand{\brightNano}{\pairs\per\second\per\mega\hertz\per\nano\watt}
\newcommand{\GammaDone}{\Gamma_{\mathrm{D}1}}
\newcommand{\GammaDtwo}{\Gamma_{\mathrm{D}2}}

\date{\today}

\begin{document}
\title{Ultrabright and narrowband intra-fiber biphoton source at ultralow pump power}

\author{Alexander Bruns}
\affiliation{Institut für Angewandte Physik, Technische Universität Darmstadt, Hochschulstraße 6, 64289 Darmstadt, Germany}
\email{alexander.bruns@tu-darmstadt.de}
\author{Chia-Yu Hsu}
\affiliation{Institut für Angewandte Physik, Technische Universität Darmstadt, Hochschulstraße 6, 64289 Darmstadt, Germany}
\affiliation{Department of Physics, National Tsing Hua University, Hsinchu 30013, Taiwan}
\author{Sergiy Stryzhenko}
\affiliation{Institut für Angewandte Physik, Technische Universität Darmstadt, Hochschulstraße 6, 64289 Darmstadt, Germany}
\affiliation{Institute of Physics, National Academy of Science of Ukraine, Nauky Avenue 46, Kyiv 03028, Ukraine}
\author{Enno Giese}
\affiliation{Institut für Angewandte Physik, Technische Universität Darmstadt, Hochschulstraße 6, 64289 Darmstadt, Germany}
\author{Leonid P. Yatsenko}
\affiliation{Institute of Physics, National Academy of Science of Ukraine, Nauky Avenue 46, Kyiv 03028, Ukraine}
\author{Ite A. Yu}
\affiliation{Department of Physics, National Tsing Hua University, Hsinchu 30013, Taiwan}
\affiliation{Center for Quantum Technology, Hsinchu 30013, Taiwan}
\author{Thomas Halfmann}
\affiliation{Institut für Angewandte Physik, Technische Universität Darmstadt, Hochschulstraße 6, 64289 Darmstadt, Germany}
\author{Thorsten Peters}
\affiliation{Institut für Angewandte Physik, Technische Universität Darmstadt, Hochschulstraße 6, 64289 Darmstadt, Germany}

\begin{abstract}
    Nonclassical photon sources of high brightness are key components of quantum communication technologies. We here demonstrate the generation of narrowband, nonclassical photon pairs by employing spontaneous four-wave mixing in an optically-dense ensemble of cold atoms within a hollow-core fiber. The brightness of our source approaches the limit of achievable generated spectral brightness at which successive photon pairs start to overlap in time. For a generated spectral brightness per pump power of up to \SI{2e9}{\brightMilli} we observe nonclassical correlations at pump powers below \SI{100}{\nano\watt} and a narrow bandwidth of $2\pi\times$ \SI{6.5}{\mega\hertz}. 
    In this regime we demonstrate that our source can be used as a heralded single-photon source. 
    By further increasing the brightness we enter the regime where successive photon pairs start to overlap in time and the cross-correlation approaches a limit corresponding to thermal statistics.
	Our approach of combining the advantages of atomic ensembles and waveguide environments is an important step towards photonic quantum networks of ensemble-based elements.
\end{abstract}
\maketitle
\section{Introduction\label{sec:Intro}}
Nonclassical photon sources are of paramount importance for optical quantum communication \cite{sangouard_quantum_2011,slussarenko_photonic_2019}. Key requirements for these sources are high single-photon purity and high brightness, which are typically difficult to achieve simultaneously. If these photon sources are to be interfaced with atomic or solid-state quantum memories, spectrally narrow single photons in the \si{\mega\hertz} regime are additionally required to match the relevant atomic transitions \cite{eisaman_invited_2011}. In the context of quantum networks \cite{kimble_quantum_2008} and moving towards real-life applications, the integrability of the technology into optical waveguides becomes crucial \cite{wang_integrated_2021}. This, e.g., has the advantage of better scalability, lower pump powers, and improved efficiency due to better mode-matching as compared to free-space setups.

While deterministic sources include the timing information by design, probabilistic sources require an additional heralding mechanism. It is commonly realized by implementing a probabilistic source of correlated photon pairs and using one of the photons as a herald to obtain timing information about the second photon \cite{eisaman_invited_2011}. Such correlated photon pairs are typically generated by nonlinear optical processes such as spontaneous parametric down-conversion (SPDC) or spontaneous four-wave mixing (SFWM). Both of these processes can provide a high spectral biphoton brightness at narrow bandwidths either due to the inherent frequency filtering of SFWM or by employing an additional cavity. The highest generated spectral brightnesses (GSB) reported so far in excess of \SI{1e5}{\bright} were achieved using a variety of different experimental systems. These are waveguides combined with cavities \cite{luo_direct_2015,steiner_ultrabright_2021}, a bulk crystal inside a cavity \cite{tsai_ultrabright_2018}, and a room-temperature atomic ensemble  \cite{chen_room-temperature_2022}.
The bandwidths of these biphoton sources ranged from sub-MHz \cite{chen_room-temperature_2022} to around \SI{100}{\mega\hertz} \cite{steiner_ultrabright_2021} and the generated spectral brightness per pump power (GSBP) ranged from \SI{3e4}{\brightMilli} \cite{luo_direct_2015} to \SI{2e8}{\brightMilli} \cite{steiner_ultrabright_2021}. While each system and technique has its own advantages and it is therefore difficult to compare the values achieved, they put our results into context.

In the following, we report on the first implementation of a biphoton source using SFWM in an ensemble of cold rubidium atoms loaded into a hollow-core fiber. To avoid collisions of the cold atoms with the fiber wall, thereby reducing the coherence time, the atoms are guided by an optical dipole trap. By using cold atoms we achieve an order of magnitude lower biphoton bandwidth as compared to fibers filled with warm gases \cite{cordier_raman-free_2020,lopez-huidobro_fiber-based_2021}. This was first shown by Corzo \textit{et al.}, who interfaced cold atoms with a nanofiber to generate narrowband single photons on-demand \cite{corzo_waveguide-coupled_2019}. However, using a hollow-core fiber instead of a nanofiber has the benefit that all light fields, including the pump, are guided in the same optical mode. Thus, we obtain intrinsically optimal mode-matching as well as strong light-atom coupling at orders of magnitude lower pump powers. This results in a GSBP of up to \SI{2e9}{\brightMilli}, which is a 10-fold increase over the previous record \cite{steiner_ultrabright_2021} at 100-fold reduced pump power and 10-fold lower bandwidth of $2\pi\times$ \SI{6.5}{\mega\hertz}, which is directly compatible with atomic quantum memories. Moreover, we show that by tuning the brightness of our source even higher, we reach the ultimate achievable limit of spectral brightness, at which successive photon pairs start to overlap in time. In this regime, the cross-correlation approaches a limit expected for thermal statistics. 
To characterize our photon source, we present thorough measurements on the cross- and auto-correlations, bandwidth, and GSB of the photon pairs. Where possible, we compare our experimental results to theoretical simulations for a quantitative analysis. Due to the high GSB, narrow bandwidth, and waveguide-coupled single-mode emission, possible applications of our biphoton source include integration into photonic quantum networks \cite{kimble_quantum_2008} using atomic ensembles as building blocks \cite{sangouard_quantum_2011}, quantum key distribution via satellite links \cite{bedington_progress_2017}, and photonic quantum metrology \cite{polino_photonic_2020}.

\begin{figure*}[t]
    \centering
    \includegraphics[width=\textwidth]{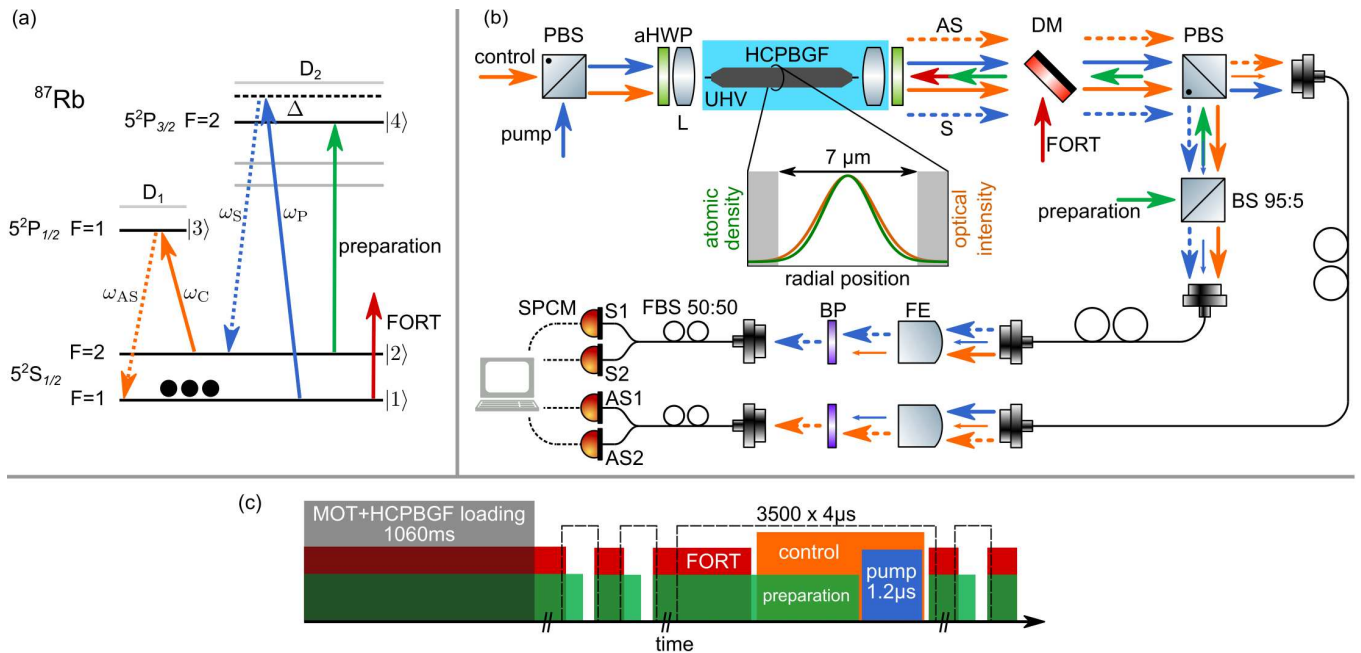}
    \caption{(a) Double-$\Lambda$ system for SFWM in ${}^{87}\text{Rb}$. The excited states decay at rates $\GammaDone=2\pi\times$ \SI{5.75}{\mega\hertz} and $\GammaDtwo=2\pi\times$ \SI{6.06}{\mega\hertz}. We measured the ground-state decoherence rate (via EIT light storage similar to \cite{peters_single-photon-level_2020}) to be $\gamma_{12}=\SI{0.057}{\GammaDone}$. For all material constants we use the values given in \cite{steck_rubidium_2021}.  (b) Simplified experimental setup. Solid/dotted arrows depict applied/generated light fields. Colors match those in the level scheme. Small arrows indicate optical noise before sufficient filtering. (P/F)BS: (polarizing/fiber) beamsplitters, aHWP: achromatic half-wave-plate, L: lens, UHV:  ultra-high vacuum, DM: dichroic mirror, FE: filter etalons, BP: optical bandpass filter. Solid lines represent optical single-mode fibers and dotted lines electronic signals.  (c) Simplified temporal sequence.}
    \label{fig:Setup}
\end{figure*}

\section{Experimental Methods \label{sec:Setup}}
Our SFWM coupling scheme for photon-pair generation using ${}^{87}\text{Rb}$ atoms is shown in figure \ref{fig:Setup}(a). 
As we do not intentionally lift the degeneracy of the Zeeman states, only the hyperfine states are shown. A pump beam (frequency $\omega_P$) with power $P$ on the D$_2$ line, blue-detuned by $\Delta=\SI{53}{\GammaDtwo}$ from the transition $\ket{1}\leftrightarrow\ket{4}$ to minimize optical pumping, generates Stokes (S) photons (frequency $\omega_S$) via spontaneous Raman scattering. The collinear control beam (frequency $\omega_C$) with Rabi frequency $\Omega_C$ on the D$_1$ line is applied resonantly to the transition $\ket{2}\leftrightarrow\ket{3}$. This creates a narrow transparency window via electromagnetically induced transparency (EIT) \cite{fleischhauer_electromagnetically_2005} for the correlated anti-Stokes (AS) photons (frequency $\omega_{AS}$) in the SFWM process \cite{balic_generation_2005}.

The nonlinear optical medium in our experiment is an ensemble of up to \num{2.5e5} atoms at a temperature of about \SI{1}{\milli\kelvin}, a length of about $L=\SI{6}{\centi\meter}$ and a radial diameter of $d=\SI{3.4}{\micro\meter}$ that is located inside a hollow-core photonic bandgap fiber (\hcf) \cite{poletti_hollow-core_2013} (\textit{NKT Photonics}, HC800-02, \SI{14}{\centi\meter} long, numerical aperture $\sim\num{0.15}$, \SI{5.5}{\micro\meter} mode field $1/\mathrm{e}^2$ diameter of intensity). The process of preparing the cold atoms inside a magneto-optical trap (MOT) and guiding them into the \hcf\ using a far-off resonant trap (FORT) at a wavelength of \SI{820}{\nano\meter} was previously explained and characterized in detail \cite{blatt_one-dimensional_2014,peters_loading_2021}. For simplicity, we therefore omit here all details regarding the preparation of the atomic ensemble inside the \hcf.
 
The optical setup of the SFWM experiment is schematically shown in figure \ref{fig:Setup}(b). All laser beams are generated using home-built diode laser systems with linewidths of several hundred kHz (except of the FORT \cite{peters_loading_2021}). The timing and frequency of the laser beams are controlled by acousto-optic modulators with switching times of about \SI{100}{\nano\second}. Because all beams of SFWM are guided in a single optical mode of the \hcf, we achieve Rabi frequencies of the order of $\Gamma$ with nanowatts of optical power only. The complete experimental setup requires an optical power of about \SI{300}{\milli\watt} to operate the MOT and the FORT. Pump and control fields are orthogonally linearly polarized and combined in front of the \hcf\ using a polarizing beamsplitter (PBS). As the involved wavelengths span \SI{40}{\nano\meter} we use achromatic half-wave plates to align the polarization axes of the linearly polarized laser fields to the optical axis of the  \hcf\ at the input. This allows for achieving the highest degree of polarization at the output and thus optimum polarization filtering as required for operation at the photon level \cite{peters_single-photon-level_2020}. As confirmed by numerical simulations (see appendix \ref{sec:appendixPolSim}) of the SFWM process including the Zeeman structure and laser field polarizations, the generated S (AS) field is orthogonally polarized to the pump (control) field. Thus, the two generated fields can be separated by a PBS which also serves as a first filter stage. The optical noise originating from residual pump (control) light in the S (AS) channel separated by \SI{6.8}{\giga\hertz} is attenuated by up to \SI{40}{\decibel}. After polarization filtering the photons are coupled into single-mode fibers which serve as spatial filters to isolate the \hcf 's optical mode. As a second filtering stage we use temperature-tuned monolithic etalons \cite{palittapongarnpim_note_2012,ahlrichs_monolithic_2013} with an attenuation of up to \SI{45}{\decibel} to further attenuate the optical noise in the S (AS) channel that originates from the pump (control) beam (separated by \SI{6.8}{\giga\hertz}) and the control (pump) beam (separated by \SI{15}{\nano\meter}). Finally, we employ narrowband optical filters (blocking optical depth (OD) $>4$) to further attenuate the \SI{15}{\nano\meter} separated optical noise and broadband noise originating from stray light. The laser beams not directly involved in the SFWM process, i.e., preparation beam and FORT, are aligned counter-propagating to the pump and control beams to protect the photon counting equipment and to minimize optical noise. The residual optical noise (measured without loading atoms into the \hcf) is about \SI{2800}{\per\second} (\SI{1200}{\per\second}) for the S (AS) channel.

To detect and characterize the generated photon pairs, we implement Hanbury Brown \& Twiss setups for both photonic channels by splitting the signals with 50:50 fiber beamsplitters before guiding them to single-photon counting modules (SPCM, \textit{Excelitas}, SPCM-AQRH-13). The overall detection efficiency in both channels is about \SI{8}{\percent}, including the transmission through all elements after the \hcf\ and the intrinsic quantum efficiencies of our detectors. We acquire the timings of all detected photons with time-tagging electronics (\textit{Swabian Instruments}, Timetagger 20) for further analysis. The combined timing jitter of the detection system is specified as $<$\SI{400}{\pico\second}, while the temporal dynamics of the biphoton waveform are on a timescale of a few \SI{10}{\nano\second} (see Figure 2). Hence, we neglect this jitter in the analysis of our data.

Figure \ref{fig:Setup}(c) shows the temporal sequence of the experiment. After the atoms are captured in the MOT and subsequently loaded into the \hcf, we start the experimental SFWM phase. To avoid any influence of the strong FORT on the SFWM process, we periodically switch off the FORT for periods of \SI{2.4}{\micro\second} before switching it on again for \SI{1.6}{\micro\second} to recapture the atoms and prevent collisions with the fiber wall. During each period where the FORT is off, the atoms are first prepared in the ground state $\ket{1}$ by optical pumping with the SFWM control beam and an additional orthogonally-polarized preparation beam on the D$_2$ line to address all magnetic sublevels of ground state $\ket{2}$. Note that we do not prepare the atoms in a single Zeeman state. Subsequently, while keeping the control beam on, we simultaneously apply the pump beam to drive the SFWM process for a duration of \SI{1.2}{\micro\second}. We acquire data from the SPCMs only during this SFWM phase. Due to the finite temperature of the atomic ensemble, the periodic switching of the FORT leads to a non-negligible loss of atoms over time. To characterize these changing experimental conditions, we measured time-resolved absorption spectra by probing the AS transition with a weak laser pulse. We observed a peak OD of \num{155} on the $\ket{1}\leftrightarrow\ket{3}$ transition that decays exponentially with a time constant of \SI{3.2}{\milli\second}. Therefore, we choose to repeat the experimental window \num{3500} times for a total duration of \SI{12}{\milli\second}, to include the full range of ODs available to us. After this time, basically all atoms are lost and we start with a new MOT loading cycle. Because the OD is changing over two orders of magnitude during the measurements, we also modulate the control power in a similar way to maintain similar EIT conditions over a larger range of measurement windows. We can then post-select the range of experimental windows (or ODs) in the evaluation. Narrow ranges lead to sharply defined experimental conditions, while wider ranges increase the detection rates at the cost of averaging over changing conditions. The duty cycle, i.e., the ratio of SFWM data acquisition phases to the total duration of the experiment depends on the post-selected parameter ranges and is typically around 1:2000. The generation rates given throughout this paper are corrected for the respective duty cycle.

\section{Experimental Results \label{sec:Results}}
By implementing the experiment as described in the previous section, we generate correlated S-AS photon pairs for a wide range of the experimentally controllable parameters OD and pump power. This allows for tuning the spectral brightness and cross-correlation of our photon-pair source, which we will discuss later in more detail.

\begin{figure*}[tb]
    \centering
    \includegraphics[width=\textwidth ]{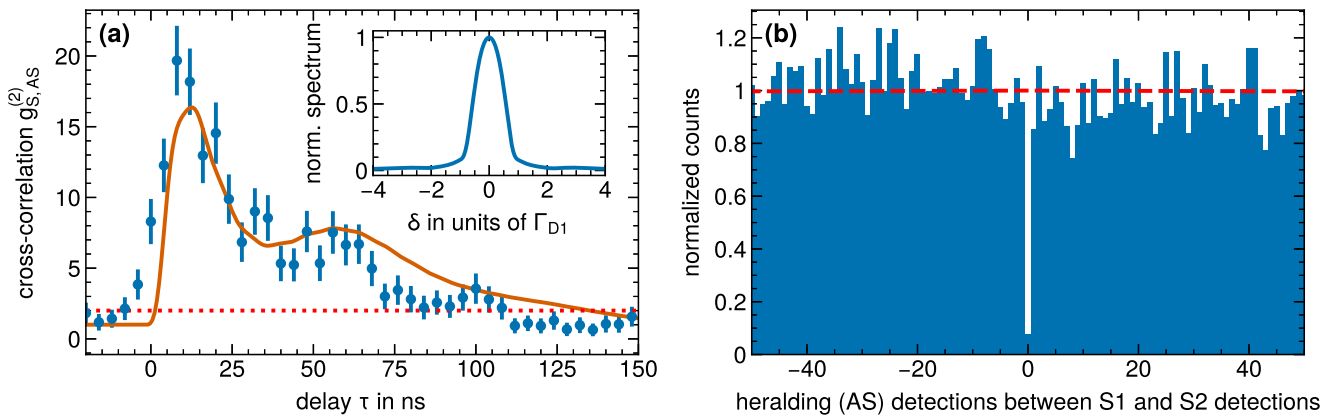}
    \caption{(a) Exemplary temporal biphoton waveform. Cross-correlation \gTwo{S,AS} as a function of the delay $\tau$ between AS and S detection events. The error bars depict the propagated Poissonian error of the photon counts. The red dotted line indicates the upper limit of the classical regime \gTwo{S,AS}$=2$. We apply a pump power of $P=\SI{14}{\nano\watt}$ and a control power of \SI{17}{\nano\watt} corresponding to a Rabi frequencies of $\Omega_P=\SI{3.1}{\GammaDone}$ and $\Omega_C=\SI{2.8}{\GammaDone}$. We include measurements in the range of $OD=40-80$ to obtain good statistics within an effective (i.e., duty-cycle corrected) integration time of \SI{3.5}{\second}. The bin size is \SI{4}{\nano\second}. The solid line shows the results of a numerical simulation of equation \eqref{eqBiphotonWaveform} using parameters $OD={15}$, $\Omega_C=\SI{2.8}{\GammaDone}$ and $P=\SI{14}{\nano\watt}$ and is vertically scaled to fit the datapoints. A fixed factor of \gTwo{S,AS}$=1$ is added to the simulation to account for the uncorrelated background.  The inset shows the normalized biphoton spectrum extracted from this simulation. (b) Measured heralded auto-correlation \gTwo{S,S|AS} of the S photon conditioned on the detection of an AS photon. Dataset is the same as in (a). See the main text for details on the experimental method.}
    \label{fig:ResultBiphotonQuality}
\end{figure*}

\begin{figure}[bt]
    \centering
    \includegraphics[width=\columnwidth]{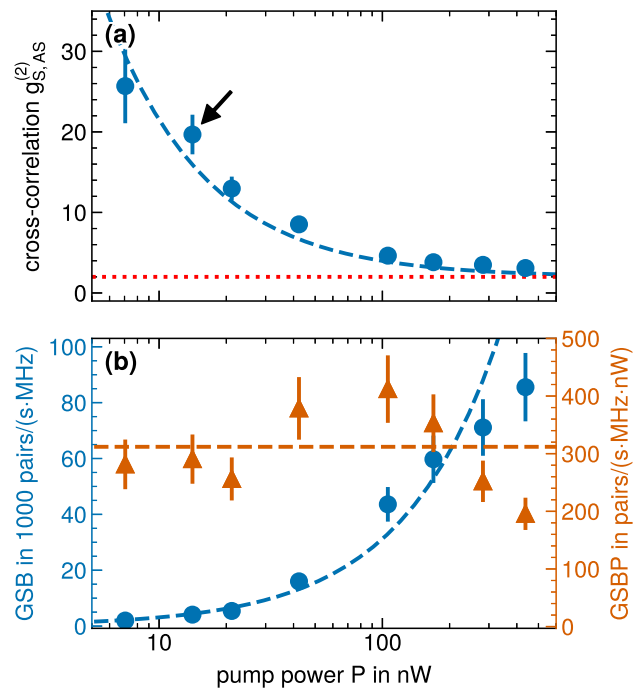}
    \caption{Photon pair properties versus pump power. The experimental parameters are $OD=40-80$ and the control Rabi frequency $\Omega_C=\SI{2.8}{\GammaDone}$. The effective integration times range from \SIrange{0.6}{3.5}{\second}. (a) Measured peak cross-correlation \gTwo{S,AS} (blue circles). The data point indicated by an arrow corresponds to the measurement in figure \ref{fig:ResultBiphotonQuality}. The dotted red line visualizes the classical boundary set by the Cauchy-Schwarz inequality. The dashed blue line is a numerical model including noise contributions. The parameters are the same as for the model shown in figure \ref{fig:Result2D}(c). (b) GSB (left axis, blue circles) and  GSBP (right axis, orange triangles). The dashed blue line is a linear fit to the brightness in the low pump power regime. The dashed orange line corresponds to the same fit additionally normalized to the pump power. The error bars include the Poissonian error of the photon counts and a \SI{10}{\percent} variation in the total detection efficiencies.}
    \label{fig:ResultVsRate}
\end{figure}

First, however, we discuss the properties of the observed photon pairs at fixed experimental conditions yielding a good signal-to-noise ratio at a GSBP of about \SI{300}{\brightNano} (see discussion of figure \ref{fig:ResultVsRate}) which is comparable to the record GSBP reported in \cite{steiner_ultrabright_2021}. Figure \ref{fig:ResultBiphotonQuality} (a) shows exemplary data (blue circles) of the cross-correlation $g^{(2)}_{S,AS}(\tau)$ of the photon pairs as a function of the time delay $\tau$ between S and subsequent AS detection events. We obtained the data by normalizing the coincidences to the background of accidental coincidences. We acquired this background by correlating events from different measurement windows, hence removing any physical correlation from the data. 
Clearly, the cross-correlation exceeds the classical limit of $g^{(2)}_{S,AS}(\tau)=2$ (red dotted line) for a range of delays. The small damped oscillations visible in the waveform indicate that we work in the transition region between the damped Rabi oscillation and the group delay regimes, as defined in \cite{du_narrowband_2008,kolchin_electromagnetically-induced-transparency-based_2007}. We verified experimentally that the individual S and AS fields exhibit thermal statistics, i.e., $g^{(2)}_{S,S}(0)=g^{(2)}_{AS,AS}(0)\approx2$. To obtain sufficiently good statistics for these measurements within a feasible integration time, we had to raise the generation rate by increasing the pump power to $\gtrsim\SI{100}{\nano\watt}$.
To quantify the violation of the Cauchy-Schwarz inequality \cite{clauser_experimental_1974} we calculated $\mathcal{R}=({g^{(2)}_{S,AS}})^2/(g^{(2)}_{S,S}\,g^{(2)}_{AS,AS})$. The peak value is  $\mathcal{R}=\num{97(24)}$, which clearly violates the classical limit of $\mathcal{R}\leq 1$ by 4 standard deviations and demonstrates the nonclassical nature of the photon pairs. 
To obtain the bandwidth of the photon-pair source, we compare the experimental data shown in figure \ref{fig:ResultBiphotonQuality}(a) to a simulated waveform using the theory presented in \cite{du_narrowband_2008}. The relevant equations used are summarized in appendix \ref{sec:Theory}. First, we evaluated equation \eqref{eqBiphotonWaveform} starting with parameters determined from other measurements and subsequently optimized these parameters for the best agreement between simulation and experimental data. The features of the experimental data (blue circles) can be reproduced by the simulation (orange solid line). The discrepancy between the parameter sets of experiment and simulation might be explained by deviations of our experiment from the assumptions made in \cite{du_narrowband_2008}. There, homogeneous atomic density and pump/control intensities are assumed, which is clearly not the case inside the \hcf. Moreover, the simulation considers only population of a single Zeeman level, whereas in our case the population is initially distributed among the Zeeman levels of $\ket{F=1}$. We use the simulated waveform to extract the biphoton spectrum as shown in the inset of figure \ref{fig:ResultBiphotonQuality}(a). Here, $\delta$ is the single-photon detuning of the generated AS photon frequency components from their central frequency $\omega_{AS}$. The spectral bandwidth (FWHM) is $2\pi\times\SI{6.5}{\mega\hertz}\approx\SI{1.1}{\GammaDone}$ demonstrating compatibility of our source with other rubidium-based experiments. The characteristic timescale of the biphotons is thus \SI{24}{\nano\second}.

We now turn to interpreting the photon-pair source as a heralded single-photon source. Instead of using the S for heralding the AS detection events, as it is typically done, we use the AS to herald the S events for the following reason: As the Raman gain is not negligible compared to the FWM gain for our current parameters, uncorrelated photons can be generated which act as additional noise. This process is mainly relevant for the S channel as the pump is coupled to the populated state $\ket{1}$, whereas the control is coupled to the unpopulated state $\ket{2}$. Using the AS detection events as heralding events for the creation of single photons in the S channel, we evade the influence of Raman noise on the heralding itself. As the AS photon is delayed by the slow-light effect of EIT \cite{fleischhauer_electromagnetically_2005} and thus exits the medium after the S photon, choosing the AS as heralding photon is somewhat counter-intuitive. If necessary, however, the temporal order of S and AS detection events could be easily changed by sending the S photons through a fiber optical delay line. 
In figure \ref{fig:ResultBiphotonQuality}(b) we plot the auto-correlation of the S photons conditioned on the detection of an AS photon, \gTwo{S,S|AS}. To obtain data of sufficient quality within a reasonable acquisition time, we use the following method \cite{fasel_high-quality_2004,seri_laser-written_2018}: Any heralded S1(S2) detection event (see figure \ref{fig:Setup}) is used as a start trigger. For any subsequent S2(S1) detection event, we record the number of additional AS events $n$ between the two S events. The number of S1-S2(S2-S1) pairs with $n$ AS events in between gives the amplitude of the $n$-th ($-n$-th) bin in the resulting histogram. The zero bin thus corresponds to more than one S photon being heralded by the same AS photon. From this data we can now determine \gTwo{S,S|AS}$(0)$ as the amplitude of the zero bin normalized to the expected value that we get from a linear fit (red dashed line) to the remaining bins \cite{fasel_high-quality_2004}.  We observe anti-bunching of the heralded S photons with \gTwo{S,S|AS}$(0)=\num{0.08(1)}$, clearly violating the classical bound of \gTwo{S,S|AS}$(0)\ge 1$ \cite{grangier_experimental_1986} and confirming a non-zero projection onto the single-photon Fock state for \gTwo{S,S|AS}$(0)\le 0.5$ \cite{grunwald_effective_2019}. We calculate the heralding efficiency as the ratio of the pair generation rate and the heralding (AS) rate corrected for transmission losses and find a value of \SI{43}{\percent}. The optimal mode matching in the \hcf\ facilitates such efficient heralding. Similar values were reported using a nanofiber \cite{corzo_waveguide-coupled_2019}.

\begin{figure}[tb]
    \centering
    \includegraphics[width=\columnwidth ]{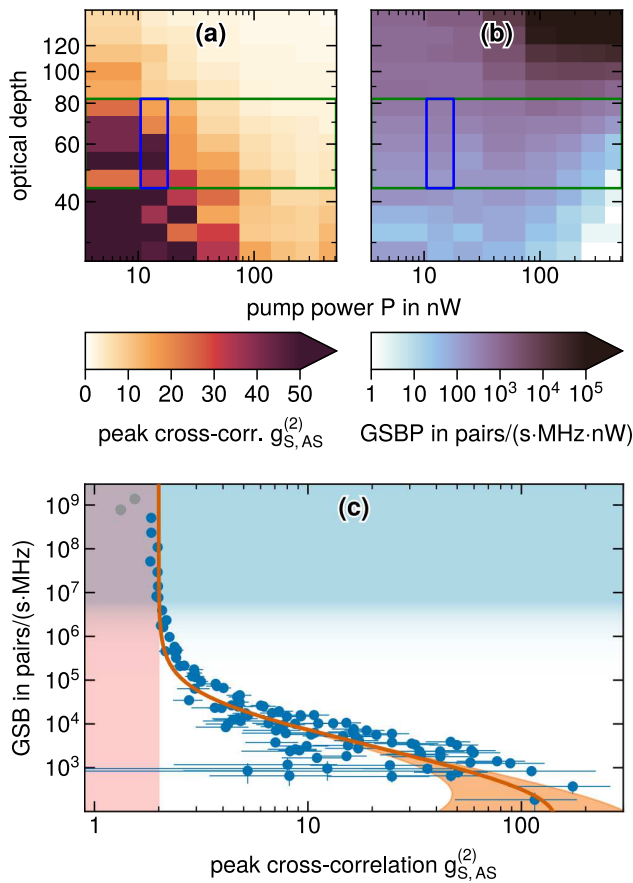}
    \caption{(a) Peak cross-correlation and (b) GSBP vs. pump power and OD. The control Rabi frequency for $OD\gtrsim 55$ is \SI{2.8}{\GammaDone}, for smaller ODs it is continuously reduced to maintain a fixed width of the EIT transmission spectrum. The blue (green) box marks the data used in figure \ref{fig:ResultBiphotonQuality} (figure \ref{fig:ResultVsRate}). The open-end color scales visualize the attainable values due to the limited measurement time. (c) GSB vs. cross-correlation \gTwo{S,AS}. The data is taken from (a) and (b). The classical regime of \gTwo{S,AS}$\leq2$ is marked by the red shaded area. The blue-shaded area marks the GBS where subsequent photon pairs overlap in time (see main text). The gradient illustrates the transition into this regime.  The error bars of the data points correspond to the propagated statistical error of the photon counts. For the grey data points the saturation of the SPCMs due to high detection rates can no longer be neglected. The solid orange line is a single-mode model described by equation \eqref{eqGSB}. The shaded orange area visualizes the influence of pump-dependent noise as explained in the main text.}
    \label{fig:Result2D}
\end{figure}

Next, we analyze the performance of our biphoton source when varying the pump power $P$ over three orders of magnitude. Figure \ref{fig:ResultVsRate}(a) shows the measured peak cross-correlation \gTwo{S,AS} (blue circles) as a function of $P$. With increasing pump power the cross-correlation, i.e., the purity of the photon pairs, reduces following the expected $1/P$ dependency. Nevertheless, we observe nonclassical correlations over the full range of pump powers. We use the model described in appendix \ref{sec:NoiseModel} to calculate the peak cross-correlation that includes the measured unconditional as well as pump power-dependent optical noise in the S and AS channel (dashed line). The model is described in more detail in the discussion of figure \ref{fig:Result2D}(c). The same parameter set determined in figure \ref{fig:Result2D}(c) is also used for the simulation shown in figure \ref{fig:ResultVsRate}(a) without any additional fitting parameter. We use a linear dependence between the generated brightness and the pump power as shown in \ref{fig:ResultVsRate}(b) (see next paragraph). Experiment and theory agree well over the whole range of pump powers.

In figure \ref{fig:ResultVsRate}(b) we show the corresponding GSB (blue circles, left hand side axis). We obtain this rate from the detected rate by correcting it for optical background, transmission losses, detection efficiencies, and the duty cycle of the experiment. The dashed blue line is a linear fit of type $\textrm{GSB}=\textrm{GSBP}\cdot P$. The fit confirms the expected linear dependence for pump powers up to \SI{200}{\nano\watt}. For higher pump powers, the spectral generation rate increases slower. We suspect that in this regime population redistribution due to optical pumping is no longer negligible. When we normalize the spectral generation rate with regard to the pump power, we obtain the GSBP. The experimental values (orange triangles, right hand side axis) range from \SIrange{200}{410}{\per\s\per\mega\hertz\per\nano\watt}. The orange dashed line represents the fitted value $\textrm{GSBP}=\SI{312(24)}{\brightNano}\approx \SI{3e8}{\brightMilli}$. In this intermediate parameter regime the GSBP of our source is comparable to the highest reported value of \SI{2e8}{\brightMilli} using a waveguide coupled to an on-chip microring cavity \cite{steiner_ultrabright_2021}, however, at a 10-fold reduced bandwidth and a 100-fold reduced pump power. This very efficient conversion of pump power into narrowband photon pairs is enabled by the intrinsically large overlap between the cold atomic ensemble and the light fields in the \hcf\ as well as the optimal mode matching between the four involved fields due to all of them being guided in a single optical mode. 

Next, we investigate the dependence of the peak cross-correlation \gTwo{S,AS} and the GSBP on the pump power as well as the OD. The results are shown in figure \ref{fig:Result2D}. We note that our system allows for an easy tuning of the OD over an order of magnitude. This, combined with a tuning of the pump power results in a tunability of the GSBP of our biphoton source by two orders of magnitude. The regimes of high generation rate but low-quality photon pairs (upper right corner at high OD and high pump power) and low generation rate with high-quality photon pairs (lower left corner at low OD and low pump power) can be clearly identified in figure \ref{fig:Result2D}(a) and (b). The parameter range of the data presented in figures \ref{fig:ResultBiphotonQuality} and \ref{fig:ResultVsRate}  lies in between these two regimes and is marked by the blue and green rectangles. The highest values for \gTwo{S,AS} are observed in the low OD and low pump power regime. When we restrict ourselves to parameters with \gTwo{S,AS}$\geq 3$, the highest observed GSBP is \SI{2e9}{\brightMilli}.  For values \gTwo{S,AS}$\gtrsim50$, the relative uncertainties become larger than \SI{30}{\percent}, due to the low number of counts during the finite measurement duration. This measurement demonstrates the versatile nature of our photon-pair source. Depending on the application requirements, the parameter regime can be easily adjusted. 

Finally, we investigate the limit of the GSB achieved with our system. To this end, we plot in figure \ref{fig:Result2D}(c) the GSB vs. the peak cross-correlation \gTwo{S,AS}. The data are taken from figures (a) and (b). Note that the grey datapoints correspond to a count rate where the saturation of the SPCMs can not be neglected anymore. As we can see, for higher spectral brightnesses the peak cross-correlation decreases and reaches a lower limit near two. We fitted the data using the model described in appendix \ref{sec:NoiseModel} that is based on a single-mode description and thermal statistics.  We start from equation \eqref{eq:NoiseModel} and write the pair number as $n=\alpha\cdot\mathrm{GSB}$ with a scaling factor $\alpha$. The orange line thus corresponds to a fitted function of the form
\begin{equation}
    \label{eqGSB}
    g^{(2)}_\text{S,AS}  = \frac{ 2 +\frac{1}{\alpha\mathrm{GSB}} + \frac{\mathcal{N}_{S}\mathcal{N}_{AS}}{T_{S}T_{AS} (\alpha\mathrm{GSB})^2}}{\left[1 + \frac{\mathcal{N}_{S}}{T_{S} \alpha\mathrm{GSB}}  \right]\left[1 + \frac{\mathcal{N}_{AS}}{T_{AS} \alpha\mathrm{GSB}}  \right]}
\end{equation}
with noise contributions $\mathcal{N}_j$ and detection efficiencies $T_j$, where we used measured values as described in appendix \ref{sec:NoiseModel}. The scaling factor is the only free parameter for our fit and was determined to be $\alpha=103$. As the dataset includes points obtained with different pump powers, the pump-dependent noise contributions also vary. This is visualized by the orange-shaded area.
Nonclassical correlations with \gTwo{S,AS}$>3$ are maintained up to a GSB of around $\SI{2e5}{\bright}$. 
Beyond this brightness we approach the ultimate obtainable limit of $\mathrm{GSB}=\pi/2\times\SI{e6}{\bright}$, where successive photon pairs start to overlap in time \cite{chen_room-temperature_2022}. The transition from temporally separated to overlapping pairs is visualized by the blue color gradient in figure \ref{fig:Result2D}(c). In this range, the peak value of $g^{(2)}_{S,AS} \to 2$ indicates that thermal statistics apply while the S and AS fields are still correlated and can be described by a single mode. A detailed discussion of this regime is beyond the scope of this work and will be addressed in future work. Nonetheless, our data clearly demonstrates that our biphoton source can be operated near the ultimate limit of generated spectral brightness with a high tunability over two orders of magnitude, narrow bandwidth, and at ultralow pump powers.

\section{Conclusion and Outlook\label{sec:Conclusion}}
We demonstrated the first nonclassical photon-pair source based on SFWM in an ensemble of cold Rubidium atoms interfaced with a \hcf. The strong confinement of atoms and light fields within the \hcf\ leads to enhanced optical nonlinearities as compared to free space setups. It results in a GSBP of up to \SI{2e9}{\brightMilli} at pump powers in the regime up to \SI{100}{\nano\watt} and for a cross-correlation \num{\geq3}. We determined the biphoton linewidth as $2\pi\times$\SI{6.5}{\mega\hertz} which is thus compatible with atomic quantum memories and similar to the linewidth achievable with free-space setups. This represents a 10-fold increase of the GSBP compared to the previous record using a microring cavity, at a 10-fold reduced bandwidth and a 100-fold reduced pump power. Furthermore, the photon pairs exhibit orthogonal polarizations, thereby allowing for a relatively simple separation without the need for elaborate filtering and separation techniques \cite{caltzidis_atomic_2021}.
We verified the nonclassical nature of the photon pairs by measuring their cross-correlation with a violation of the Cauchy-Schwarz inequality by a factor of \num{97(24)} as well as a heralded auto-correlation \num{\ll0.5}.
Moreover, by increasing the generated spectral brightness even further, we reached the regime where different photon pairs start to overlap in time. Here, we demonstrated that the cross-correlation approaches a limit corresponding to thermal statistics, i.e., S and AS photons exhibit bunching.
Our hollow-core fiber-based biphoton source therefore combines the advantages of free-space and waveguide photon sources. 

Examples of possible applications include easy integration into photonic networks \cite{kimble_quantum_2008} based on atomic ensembles \cite{sangouard_quantum_2011} due to the narrowband emission into a single transverse mode, employment in quantum key distribution satellite links \cite{bedington_progress_2017} due to suitable transmission band wavelengths of \SI{780}{\nano\meter} and \SI{795}{\nano\meter}, and as light source for photonic quantum metrology \cite{polino_photonic_2020}. Additionally, the ultralow pump power required to generate biphotons combined with integrating the \hcf\ with an atom chip \cite{keil_fifteen_2016} holds promise for a significant miniaturization of the setup. Finally, the high GSB could be applied, e.g., to increase the brightness of temporally-multiplexed photon sources \cite{meyer-scott_single-photon_2020}.

To further improve and extend our \hcf-based biphoton source one could envision the following options: First, using a ladder-type scheme as in \cite{lee_highly_2016,davidson_bright_2021} the currently dominating Raman noise could be avoided.
Second, replacing our currently pulsed loading of the \hcf\ by a continuous loading scheme, the duty cycle could be increased substantially. Combined with in-fiber magic-wavelength trapping \cite{hilton_dual-color_2019,yoon_laser-cooled_2019} photon pairs could be generated continuously without the need to periodically modulate the guiding potential. We estimate that this could enhance the detected pair rates by up to three orders of magnitude. Third, by implementing intra-fiber laser cooling \cite{wang_enhancing_2022} the ensemble temperature could be reduced thereby prolonging the coherence time, i.e., further reducing the bandwidth.
Moreover, we note that as the generation of biphotons only requires atoms loaded \textit{into} instead of \textit{through} the fiber (note that in our case only the first 6~cm of the 14~cm long fiber are filled), the output side of the \hcf\ could be spliced directly to a (polarization-maintaining) fiber \cite{thapa_splicing_2006,kristensen_low-loss_2008}, thereby enabling direct connection to a photonic network, possibly after wavelength conversion to the telecom band in a waveguide \cite{albrecht_waveguide_2014}.
Finally, a detailed investigation of the photon statistics for the highest spectral brightness, where successive photon pairs overlap in time, may permit the generation of multi-photon states.

\section*{Acknowledgments}
The authors thank the group of T. Walther for providing us with their home-built ultra-low noise laser diode driver with high modulation bandwidth. The project received funding from the European Union’s Horizon 2020 research and innovation programme under the Marie Skłodowska-Curie grant agreement No. 765075, as well as by the Deutsche Forschungsgemeinschaft (DFG, German Research Foundation) under project number 410249930.

\appendix

\section{Theoretical treatment of SFWM \label{sec:Theory}}
\subsection{Biphoton waveform \label{sec:TheoryBiphoton}}
We here summarize the basic theory to simulate the biphoton waveform and spectrum based on the work of Du \textit{et al.}  \cite{du_narrowband_2008}. We focus only on  the relevant equations that enable comparison with our experimental results in figure \ref{fig:ResultBiphotonQuality}.

The double $\Lambda$-type system is introduced in figure \ref{fig:Setup}(a) of the main text. We additionally define the single-photon detuning of the AS frequency components $\omega$ as $\delta=\omega-\omega_{AS}$. Due to the collinear setup the phase matching condition, written in terms of the wavenumbers $k_i$, reads $\Delta k=k_{AS}+k_{S}-k_{C}-k_{P}=0$. We assume that all atoms are initially prepared in ground state $\ket{1}$.

We are interested in temporal correlations of the created photon pairs as a function of the time delay $\tau$ between the paired $S$ and $AS$ photons. They are determined by the relative biphoton wavefunction
\begin{equation}
    \label{eqBiphotonWaveform}
    |\psi(\tau)|^2 = \left|
        \frac{L}{2\pi} \int \kappa(\delta) \Phi(\delta) \text{e}^{-\text{i}\delta\tau} \mathrm{d}\delta
    \right|^2
\end{equation}
with the longitudinal detuning function 
\begin{equation}
    \Phi(\delta)=\text{sinc}\left(\frac{\Delta k L}{2}\right)\text{e}^{\text{i}(k_{AS}+k_{S})L/2}
\end{equation}
and the nonlinear parametric coupling coefficient
\begin{equation}
    \kappa(\delta)=-\text{i}\frac{\sqrt{\omega_{AS}\omega_{S}}}{2c}\chi^{(3)}(\delta)E_P E_C,
\end{equation}
where $E_i$ are the electric fields of the respective transitions. The third-order nonlinear susceptibility of the AS field is given by
\begin{equation}
\begin{aligned}
    \chi^{(3)}(\delta) &= \frac{
        \mathcal{N} \mu_{13} \mu_{32} \mu_{24} \mu_{41} / (\epsilon_0 \hbar^3)
    }{
         (\Delta + \text{i} \frac{\GammaDtwo}{2})
    \left[
        |\Omega_C|^2 - 4(
            \delta + \text{i} \frac{\GammaDone}{2}
        ) (\delta+ \text{i} \gamma_{12})
    \right].
    }
\end{aligned}
\end{equation}
We introduced here the transition dipole matrix elements $\mu_{ij}$, the atomic density $\mathcal{N}=N/(L \pi (d/2)^2)$ and the control Rabi frequency $\Omega_C=\mu_{23}E_C/\hbar$.

To evaluate the detuning function $\Phi(\delta)$, we calculate the complex wave numbers of the created fields as
$k_i(\delta)=\frac{\omega_i}{c}\sqrt{1+\chi_i(\delta)}$ with the linear susceptibilities given by
\begin{align}
    \chi_S(\delta) &= \frac{
        \mathcal{N} |\mu_{24}|^2 (\delta-\text{i}\frac{\GammaDone}{2}) / (\epsilon_0\hbar)
    }{
        |\Omega_C|^2 - 4(\delta - \text{i}\frac{\GammaDone}{2}) (\delta-\text{i}\gamma_{12})
    } \times \frac{
        |\Omega_P|^2
    }{
        \Delta^2+\left(\frac{\GammaDtwo}{2}\right)^2
    }, \\
    \chi_{AS}(\delta)&=\frac{4\mathcal{N}|\mu_{13}|^2(\delta+\text{i}\gamma_{12})/(\epsilon_0\hbar)}{|\Omega_C|^2-4(\delta+\text{i}\frac{\GammaDone}{2})(\delta+\text{i}\gamma_{12})}.
\end{align}
Here, $\Omega_P=\mu_{14}E_P/\hbar$ is the pump Rabi frequency.

Using these equations allows for calculating the expected waveform $|\psi(\tau)|^2$ and the biphoton spectrum $|\kappa(\delta)\Phi(\delta)|^2$.

\subsection{Polarization-resolved spatio-temporal simulation of SFWM\label{sec:appendixPolSim}}
\input{appendix_theory.tex}

\section{
Model for the detected cross-correlation
\label{sec:NoiseModel}
}
In this section we present a simple single-mode model~\cite{rohde_modelling_2006} that incorporates loss and detector inefficiencies for the fit of the cross-correlation function $g^{(2)}_{S,AS}$.
For that, we introduce the bosonic annihilation operator
\begin{equation}
\label{eq.det_photons}
    \hat{b}_j= t_j \hat{a}_j + r_j \hat{\mu}_j
\end{equation}
of photons detected in channel $j={S,AS}$.
Here, $\hat{a}_j$ denotes the (bosonic) annihilation operator of Stokes or anti-Stokes photons generated during SFWM and $\hat{\mu}_j$ the (bosonic) annihilation operator of noise photons in channel $j$.
Imperfect detection is introduced in equation~\eqref{eq.det_photons} though a beam splitter transformation with $T_j+R_j = |t_j|^2 + |r_j|^2=1$.
Hence, $T_j$ is the efficiency of the detector including all loss in channel $j$.

If we assume no correlation between the noise and the generated photons, i.\,e.,  $\braket{\hat{a}_j^\dagger \hat{\mu}_j}=0$, we can connect the detected photon numbers
$N_j= \braket{\hat{b}_j^\dagger \hat{b}_j}=T_j n_j + \mathcal{N}_j$
to the number of generated Stokes and anti-Stokes photons $n_j=\braket{\hat{a}_j^\dagger \hat{a}_j}$.
We furthermore introduced the noise detected in channel $j$ through $\mathcal{N}_j = R_j \braket{\hat{\mu}_j^\dagger \hat{\mu}_j}$.

The detected cross-correlation function $g^{(2)}_{S,AS}=\braket{\hat{b}_{S}^\dagger\hat{b}_{AS}^\dagger \hat{b}_{S}\hat{b}_{AS}}/(N_{S} N_{AS}) $ takes the form
\begin{equation}
  g^{(2)}_{S,AS}  = \frac{\braket{\hat{a}_{S}^\dagger\hat{a}_{AS}^\dagger \hat{a}_{S}\hat{a}_{AS}}}{n_{S} n_{AS}} \frac{T_{S}n_{S} T_{AS}n_{AS}}{N_{S} N_{AS}} + \frac{\mathcal{N}_{S}\mathcal{N}_{AS}}{N_{S} N_{AS}}.
\end{equation}
Here, we have again assumed no correlation between the noise and the generated photons, i.\,e., $\braket{\hat{a}_{S}^\dagger\hat{a}_{AS}^\dagger \hat{\mu}_{S}\hat{\mu}_{AS}}=0$
and uncorrelated noise in both channels, i.\,e., $R_{S}R_{AS} \braket{\hat{\mu}_{S}^\dagger\hat{\mu}_{AS}^\dagger \hat{\mu}_{S}\hat{\mu}_{AS}}= \mathcal{N}_{S}\mathcal{N}_{AS}$.

If there is perfect correlation between Stokes and anti-Stokes fields, the number of photons generated in each channel corresponds to the number of generated pairs $n=n_{S}=n_{AS}$.
We have verified that both the Stokes and anti-Stokes fields exhibit close to thermal statistics, which directly implies $\braket{\hat{a}_{S}^\dagger\hat{a}_{AS}^\dagger \hat{a}_{S}\hat{a}_{AS}}/(n_{S} n_{AS}) = 2 + 1/n$ for perfectly correlated fields.
In this case, the cross-correlation function takes the form
\begin{equation}
    \label{eq:NoiseModel}
    g^{(2)}_\text{S,AS}  = \frac{ 2 +\frac{1}{n} + \frac{\mathcal{N}_{S}\mathcal{N}_{AS}}{T_{S}T_{AS} n^2}}{\left[1 + \frac{\mathcal{N}_{S}}{T_{S} n}  \right]\left[1 + \frac{\mathcal{N}_{AS}}{T_{AS} n}  \right]} 
\end{equation}
and reduces to the ideal case for negligible noise. For dominant noise, however, the cross-correlation approaches the limit of $1+1/(n+\mathcal{N})$.
Note that in a multi-mode model the number $2$ has to be replaced by $1+1/M$, where $M$ is the number of detected modes~\cite{ivanova_multiphoton_2006}. 

To determine the noise for the fits in main body of the paper, we make the ansatz
$\mathcal{N}_j= [r_j^0+r_j^0(P)+r_j(P)]\tau_c$, where $r_i^0$ is the noise rate with no atoms loaded in the \hcf\ and no pump beam present, i.e., optical noise from stray light, residual control light and the dark counts of the SPCMs.
The additional optical noise caused by the pump beam itself, but still measured without atoms present in the \hcf, is denoted by $ r_j^0(P)$.
The remaining contribution $r_j(P)$ accounts for the noise originating from the cold atoms and it is dominated by Raman noise and residual optical pumping, but is neglected in the remainder of our analysis.
We measured the contributions without atoms to be $r_S^0=\SI{2800}{\per\second}$, $r_{AS}^0=\SI{1200}{\per\second}$,  $r_S^0(P)=\SI{84}{\per\second\per\nano\watt}\cdot P$ and $r_{AS}^0(P)=\SI{7.6}{\per\second\per\nano\watt}\cdot P$.
The characteristic time scale $\tau_c = \SI{24}{ns}$ is assumed to be of the order of the temporal duration of a biphoton, obtained from the data in figure \ref{fig:ResultBiphotonQuality}(a).
Moreover, we use $T_{S}= T_{AS}= \num{0.08}$ in accordance with the measured transmission losses and specified detector efficiencies.

\bibliographystyle{apsrev4-1}
\bibliography{BetterBibTeX}

\end{document}

%% file: appendix_theory.tex
In this section we describe the procedure for numerical simulation of the SFWM process taking into account a total of 16 magnetic sublevels of the involved hyperfine states as well as light field polarizations. This simulation served to confirm the polarizations of the generated S and AS fields for varying polarization configurations ($\parallel$ and $\perp$) of the pump and control fields. This was necessary as we are currently unable to prepare the population in a single Zeeman state and use arbitrary polarizations configurations as in free-space setups due to the \hcf s birefringence, which requires the use of linear polarizations.

We assume classical fields, use a plane-wave approximation and describe the atom-field interactions using a density matrix approach in rotating wave approximation.
We simulate the process of SFWM  by solving a system of partial differential equations
which represent light propagation and time evolution of the density matrix operator $\hat{\rho}$ with appropriate random initial conditions.

Denoting the set of magnetic sublevels of the ground states $\ket{1}$ and $\ket{2}$  by the index $g$ and a similar set of magnetic sublevels of the excited states by the index $e$ we can write the density matrix equations in the short matrix form.
In this notation all quantum operators are spelled as block matrices:
\begin{equation}
	\hat{O} = \begin{bmatrix}
		O_{gg} & O_{ge} \\
		O_{eg} & O_{ee} \\
	\end{bmatrix}.
\end{equation}

Then the propagation equations read
\begin{equation}
	\frac{\partial E_m(z, t)}{\partial z}
	= -4\pi i k_m \mathcal{N} \operatorname{tr}[ \hat{\mu}^m_{ge} \cdot \rho_{eg}(z, t) ].
	\label{eq:sim_propagation}
\end{equation}
Here $E_m$ is the complex amplitude of the electric field in mode $m$, $k_m$
is its wavenumber, and $\hat{\mu}^m$ is the mode's dipole moment operator
where the matrix elements are taken from \cite{steck_rubidium_2021}.
The mode index $m$  runs across all fields we take into account
(pump, control, Stokes, and anti-Stokes), as well as their respective polarizations,
where we use $\sigma^\pm$ as a polarization basis.

The time evolution matrix equations for the density matrix read
(here and below we assume
$\Gamma_\mathrm{D1} \simeq \Gamma_\mathrm{D2} \simeq \Gamma$)
\begin{widetext}
\begin{equation}
\begin{aligned}
	\frac{\partial \rho_{gg}}{\partial t} &=
	i [\rho_{gg}, \Delta_g] + \frac{i}{2} (
		\rho_{ge} \cdot \Omega_{eg} - \Omega_{ge} \cdot \rho_{eg}
	) + \Gamma \, R_g \circ C_{ge} \cdot \rho_{ee} \cdot C_{eg}
	- \gamma \circ \rho_{gg}, \\
	\frac{\partial \rho_{eg}}{\partial t} &= i (
		\rho_{eg} \cdot \Delta_g - \Delta_e \cdot \rho_{eg}
	) + \frac{i}{2} (
		\rho_{ee} \cdot \Omega_{eg} - \Omega_{eg} \cdot \rho_{gg}
	) - \frac{\Gamma}{2} (C_{eg} \cdot C_{ge}) \cdot \rho_{eg},\\
	\frac{\partial \rho_{ee}}{\partial t} &=
	i [\rho_{ee}, \Delta_e] + \frac{i}{2} (
		\rho_{eg} \cdot \Omega_{ge} - \Omega_{eg} \cdot \rho_{ge}
	) - \frac{\Gamma}{2} \{ C_{eg} \cdot C_{ge}, \rho_{ee} \}. \\
\end{aligned}
\label{eq:sim_evolution}
\end{equation}
\end{widetext}
Here, $\Delta_g$ and $\Delta_e$ are diagonal matrices with the summed detunings
of the considered modes on the diagonal.
$\Omega_{ge} = \Omega_{eg}^\dag$ are submatrices of the Rabi frequency operator defined by
\begin{equation}
	\hbar \hat{\Omega} = \sum_m E_m \hat{\mu}^m.
\end{equation}
The matrices $C_{ge} = C_{eg}^\dag$ contain the Clebsch-Gordan coefficients as
follows:
\begin{equation}
	\Bra{\begin{matrix}
		F_e \\
		M_e \\
	\end{matrix}} C_{eg} \Ket{\begin{matrix}
		F_g \\
		M_g \\
	\end{matrix}}
	= \braket{F_e, M_e | F_g, M_g; 1, M_e-M_g}.
\end{equation}
$R_g$ is the Kronecker delta of the $F$ quantum number:
\begin{equation}
	\Bra{\begin{matrix}
		F_1 \\
		M_1 \\
	\end{matrix}} R_g \Ket{\begin{matrix}
		F_2 \\
		M_2 \\
	\end{matrix}} = \delta_{F_1 F_2}
\end{equation}
and the operation $\circ$ stands for the element-wise multiplication of matrices.
This way, $R_g$ prevents coherence between $\ket{1}$ and $\ket{2}$ being generated
when population is transferred from the excited to the ground levels.

The matrix $\gamma$ describes decoherence effects in the ground states:
\begin{equation}
	\Bra{\begin{matrix}
		F_1 \\
		M_1 \\
	\end{matrix}}\gamma \Ket{\begin{matrix}
		F_2 \\
		M_2 \\
	\end{matrix}} = \begin{cases}
		0 \text{ if } F_1 = F_2 \text{ and } M_1 = M_2, \\
		\gamma_{12} \text{ otherwise.} \\
	\end{cases} 
\end{equation}

To imitate the vacuum fluctuations we set the Stokes field at the
beginning of the fiber to a random value with correlation time  
$1/\Gamma$ in the following way:
\begin{equation}
\begin{aligned}
    E_{S\pm}(z=0, t=0) &= 0, \\
    E_{S\pm}(z=0, t + \delta t)
    &= E_{S\pm}(z=0, t) e^{-\frac{\Gamma_\mathrm{D2}}{2} \, \delta t} \\
    &\quad + \nu_{\pm} E_\mathrm{vacuum} \sqrt{1 - e^{-\Gamma_\mathrm{D2} \, \delta t}},
\end{aligned}
\label{initial}
\end{equation}
where $\delta t \ll 1/\Gamma$ is the time step,
$\nu_{\pm}$ is a random complex number from the  normalized
complex Gaussian distribution, and the field
$E_\mathrm{vacuum}$ corresponds to the Rabi frequency of
$\sim 10^{-5} \Gamma$.
The sign $\pm$ in the index points out that we use these
initial conditions both for the left and  the right
polarization components.

The time evolution equation (\ref{eq:sim_evolution}) with initial conditions (\ref{initial}) were solved numerically
using the Kutta-Merson method.
On each Kutta-Merson step we solved the propagation equation
(\ref{eq:sim_propagation}) by integrating its right-hand side using
Simpson's rule\footnote{L. V. Blake, \textit{A Modified Simpson’s Rule and Fortran Subroutine for Cumulative Numerical Integration of a Function Defined by Data Points}, Tech. Rep. (Defense Technical Information Center, Fort Belvoir, VA, 1971)}.

The results of these simulations confirmed that the generated S and AS fields exhibit orthogonal linear polarizations with respect to the injected pump and control fields of also linear orthogonal polarizations. Therefore it was experimentally possible to use polarization filtering (see figure \ref{fig:Setup}(b)) in addition to frequency filtering and obtain a sufficiently high extinction ratio for the strong collinear pump and control fields.